\documentclass[twocolumn,
 reprint,
 amsmath,amssymb,
 aps,
prl,
]{revtex4-2}

\usepackage{graphicx}
\usepackage{dcolumn}
\usepackage{bm}

\usepackage[left]{lineno}
\usepackage[colorlinks=true, citecolor=blue, linkcolor=black, urlcolor=blue]{hyperref}
\usepackage{xcolor}
\usepackage[T1]{fontenc}
\begin{document}

\preprint{APS/123-QED}

\title{Probing the production mechanisms of virtual photons at RHIC}

\collaboration{STAR Collaboration}

\date{\today}

\begin{abstract}
We report the measurement of direct virtual photon ($\gamma^*_\text{dir} \rightarrow e^+e^-$) in four Au+Au collision energies at $\sqrt{s_\text{NN}}$ = 14.6--54.4 GeV. The $\gamma^*_\text{dir}$ yields scale with charged-particle multiplicity ($(dN_{\mathrm{ch}}/d\eta)^{\alpha}$) as $\alpha=1.43 \pm 0.04 \text{(stat.)} \pm 0.03\text{(sys.)}$, independent of collision energies and centralities, exceeding hadronic expectations and indicating notable contribution from QGP phase. At $\sqrt{s_{\rm NN}}=$ 54.4 GeV, the direct virtual photon mass distributions are further studied over $0.1 < M_{ee} < 1.1$~GeV/$c^2$, and are found to evolve from exponential to power-law form with increasing $p_\text{T}$, signaling a shift from "blackbody" radiation to emission governed by Kroll–Wada formalism.
\end{abstract}

\maketitle

In ultra-relativistic heavy-ion collisions, the extreme temperatures and energy densities achieved can drive strongly interacting matter from a hadronic state to a deconfined phase called the quark-gluon plasma (QGP)~\cite{STAR:2005gfr,BRAHMS:2004adc,BACK200528,PHENIX:2004vcz,Chen:2026gka}, providing unique way to explore the dynamics and phase structure of Quantum Chromodynamics (QCD) under extreme conditions. 
As penetrating electromagnetic probes, direct photons, defined as photons that do not originate from late-stage, long-lived hadron decays, traverse the medium with negligible strong interactions and probe all stages of the collision~\cite{Andronic:2017pug}. In selected kinematic ranges, the direct photon measurements are enhanced in sensitivity to different thermal properties of the QGP, as well as to its space-time evolution and initial temperature~\cite{Shen:2013vja,Blau:2023bvi}. 
Prompt photons arise from initial hard partonic scatterings with a large momentum transfer in the early stage of the collision. These photons typically carry high transverse momentum and dominate the direct photon yield at large $p_\text{T}$~\cite{David:2019wpt}. Therefore, the yield of these photons scales with the number of binary nucleon-nucleon collisions ($N_\text{coll}$)~\cite{PHENIX:2014nkk}. As such, they serve as a baseline for studying partonic dynamics and testing the validity of perturbative QCD theory~\cite{PHENIX:2012krx}. In contrast, thermal photons are  predominantly emitted from the QGP and hadronic phases. They are usually observed at low $p_\text{T}$ and are sensitive to the energy density and collective motion of the QGP medium~\cite{David:2019wpt}. 
However, thermal photon radiation is significantly blue-shifted due to the strong radial flow of the fireball expansion~\cite{Shen:2013vja}. Consequently, only an effective temperature can be extracted from the  $p_\text{T}$ spectra, with more direct interpretation requiring  theoretical modeling.

Direct virtual photons refer to virtual photons that do not originate from hadronic decays. They are experimentally reconstructed through their internal conversion into $e^+e^-$ pairs. In the kinematic limit $p_{\text{T},\gamma^*_\text{dir}}\gg M_{ee}$, the Kroll–Wada formalism relates the yields of low-mass direct virtual photons to those of real direct photons~\cite{PHENIX:2009gyd,Kroll:1955zu}. Consequently, measurements of low-mass $e^+e^-$ pairs provide access to the direct photon yield, including contributions from different production mechanisms such as prompt and thermal radiation.
Direct photon production at RHIC was first measured by the PHENIX collaboration in Au+Au collisions~\cite{PHENIX:2008uif}. Over the past decade, measurements of direct photons have been reported by the STAR~\cite{STAR:2016use}, PHENIX~\cite{PHENIX:2022rsx}, and ALICE~\cite{Sahlmuller:2015mfq} collaborations. However, many significant phenomena still require further investigation. For example, it has been proposed that a universal scaling of the yield with charged-particle multiplicity. would suggest that the thermal contribution to the yields is strongly correlated with system size. Theoretical modeling predicts a power-law scaling $(dN_\text{ch}/d\eta)^\alpha$ with $\alpha\sim1.8$ from a QGP medium and 1.2 from a hadronic one~\cite{Shen:2013vja}. However, PHENIX reported a substantially smaller exponent, $\alpha\sim1.1$~\cite{PHENIX:2022rsx}. The discrepancy between the reported STAR and PHENIX yield measurements~\cite{STAR:2016use,PHENIX:2022rsx}, together with the unexpectedly large direct photon elliptic flow~\cite{PHENIX:2011oxq,Sas:2018oxv}, further complicates their theoretical interpretation~\cite{vanHees:2011vb,Sun:2024vsz} and motivates a deeper investigation into the production mechanisms of direct photons and dileptons.

Direct photons originate from distinct emission regimes characterized by their production sources and kinematic scales. At sufficiently hard scales, prompt photons arise from initial non-thermal partonic scatterings described within perturbative QCD~\cite{Peitzmann:2001mz,David:2019wpt}. Thermal photons are emitted from the locally thermalized medium, while their microscopic partonic emission rates can also be perturbatively calculable at sufficiently hard scales. Toward softer scales, medium and non-perturbative effects become increasingly important. Thus, thermal versus prompt characterizes the production source, whereas perturbative versus non-perturbative describes the underlying interactions.
The mass distribution of dileptons from low-$p_\mathrm{T}$ virtual photons produced by “blackbody” thermal radiation is expected to follow a blue-shifted exponential shape~\cite{Shen:2013vja}. At higher $p_\mathrm{T}$ ($\gtrsim1$ GeV/$c$), measurements show a power-law behavior characteristic of perturbative production~\cite{PHENIX:2009gyd,Kroll:1955zu}. These distinct spectral behaviors motivate studying their transition to probe the production mechanisms of virtual photons in heavy-ion collisions.



In this letter, we report measurements of direct virtual photons using the internal conversion method ($\gamma^{*}_\text{dir}\rightarrow e^{+}e^{-}$) and investigate the production mechanisms of virtual photons in heavy-ion collisions (details in supplementary material)~\cite{STAR:2015zal,STAR:2015zal}. The Au+Au collision data used in the direct virtual photon measurements at $\sqrt{s_{\mathrm{NN}}}$ = 14.6, 19.6, 27, and 54.4 GeV were collected with the Solenoidal Tracker at RHIC (STAR) detector~\cite{STAR:2002eio} during RHIC BES-II~\cite{Chen:2024aom}.The specific results include: (a) direct virtual photon transverse momentum spectra and integrated yields in the range of $1<p_\text{T}<3$ GeV/$c$; (b) direct virtual photon mass ($M_{ee}$) distributions in different $p_\text{T}$ intervals; and (c) excess dielectron yields normalized by mid-rapidity particle multiplicity for 0.13 $< M_{ee} < $ 0.40 GeV/$c^{2}$. 


The measured dielectron yields are compared with a hadronic cocktail simulations (CKT) including contributions from $\pi^{0}$, $\eta$, and $\eta'$ Dalitz decays, vector meson decays ($\omega$, $\phi$), and semi-leptonic decays of heavy-flavor hadrons, generated with a Toy Monte Carlo (ToyMC) simulation~\cite{STAR:2012dzw}. In particular, the $\eta$ and $\pi^{0}$ spectra are parameterized using Tsallis Blast-Wave (TBW) functions~\cite{Chen:2020zuw}, with parameters constrained by existing data~\cite{STAR:2019vcp,STAR:2017sal,Dhamija:2024cfh}. The normalization of the $\eta$ meson is constrained using global $\eta/\pi^{0}$ data~\cite{PHENIX:2006avo,Agakichiev:1998ign,WA70:1988xpf,ALICE:2018vhm,Ren:2021pzi} and is fixed to $0.470 \pm 0.017$ at $p_{\rm{T}} = 5$~GeV/$c$~\cite{PHENIX:2022rsx}, where the effect of collective flow is negligible.

\begin{figure}
\includegraphics[width=0.47\textwidth]{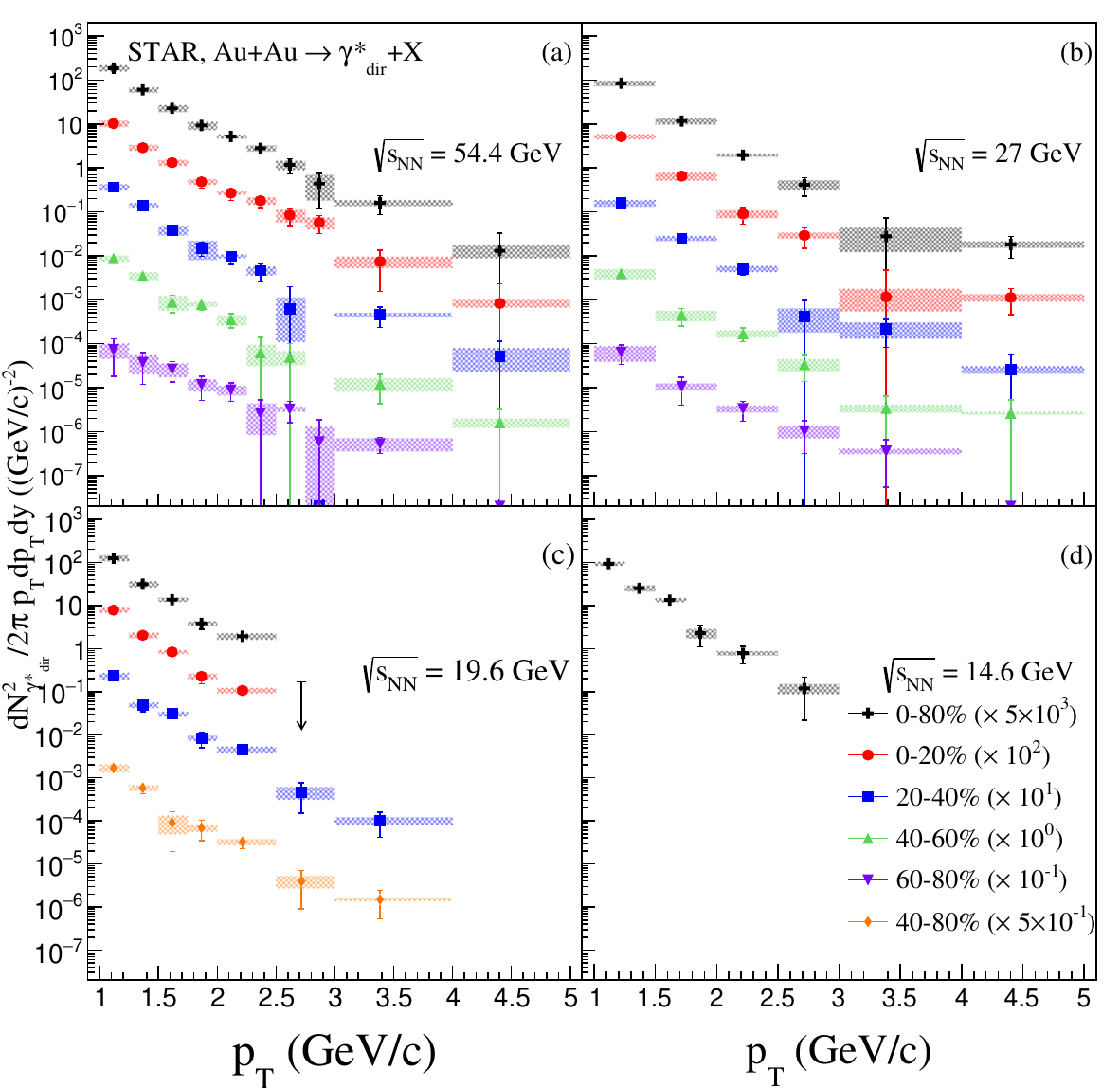}
\caption{\label{fig:DirectPhotonPtSepctra}Direct virtual photon $p_\text{T}$ spectra in Au+Au collisions at $\sqrt{s_\text{NN}}$ = 14.6, 19.6, 27, and 54.4 GeV in different centralitie. The vertical bars and boxes around data points represent the statistical and systematic uncertainties, respectively}
\end{figure}

Direct virtual photons are identified via the internal conversion technique in the kinematic regime where $p_{\text{T},\gamma^{*}_\text{dir}} \gg M_{\gamma^{*}_\text{dir}}$. In this approach, $e^{+}e^{-}$ pairs originating from direct virtual photons are reconstructed and described with a two-component fit function in the mass interval $[0.10, 0.28]$~GeV/$c^{2}$~\cite{PHENIX:2009gyd,STAR:2016use}. The invariant yield of direct virtual photons as a function of $p_{\text{T}}$ is obtained and presented in Fig.~\ref{fig:DirectPhotonPtSepctra}, with systematic uncertainty dominated by the choice of fit range and the determination of the $\eta/\pi^{0}$ production cross section.

\begin{figure}[b]
\includegraphics[width=0.47\textwidth]{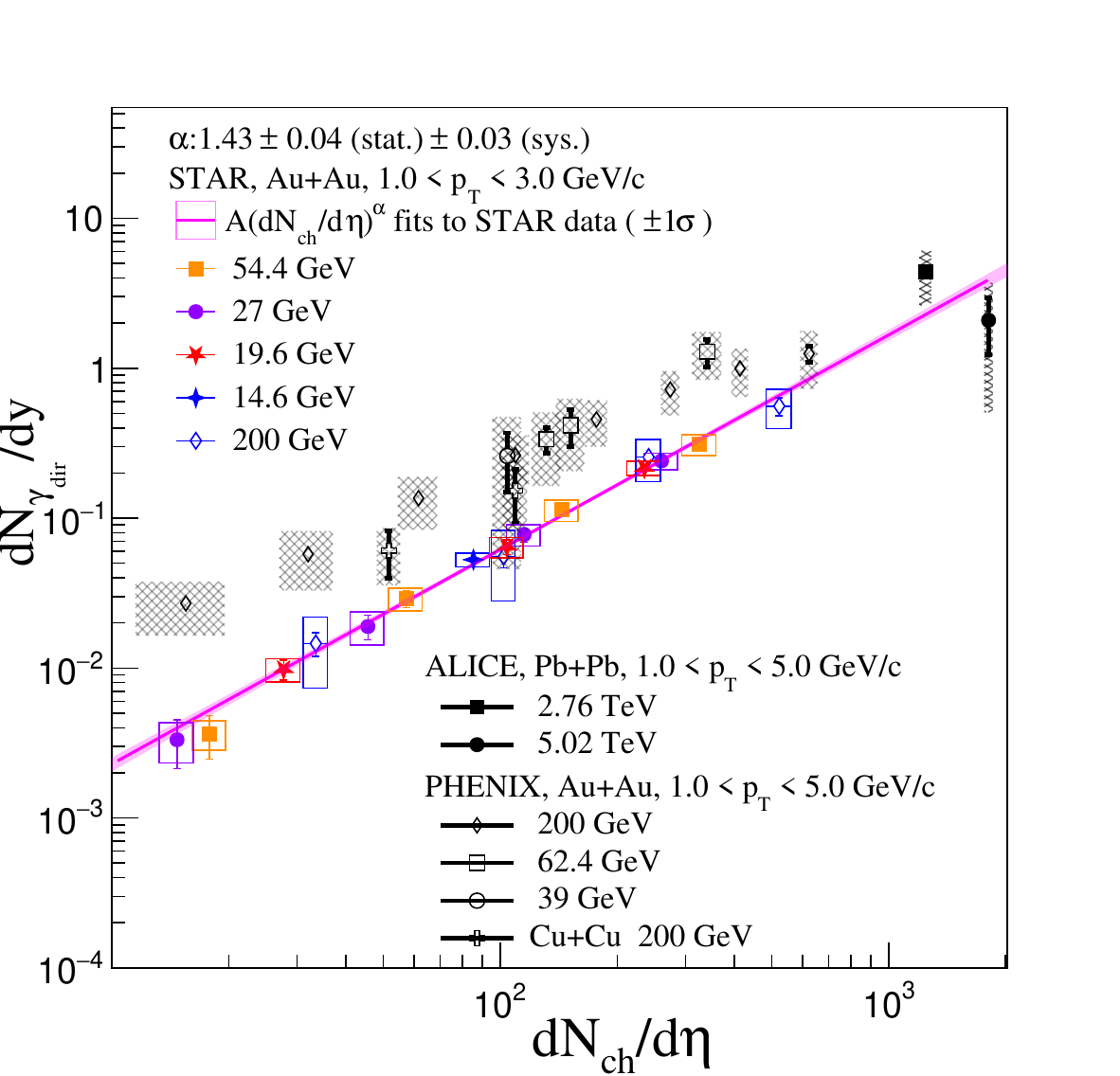}
\caption{\label{fig:DirectVirtualYield} Direct photon yields versus charged-particle multiplicity in Au+Au collisions at $\sqrt{s_\text{NN}}$ = 14.6, 19.6, 27, and 54.4 GeV. The $dN_\text{ch}/d\eta$ in each collision energy corresponds to the centrality classes used in Fig. \ref{fig:DirectPhotonPtSepctra}, respectively. Solid points denote STAR data ($1.0<p_{\mathrm{T}}<3.0$ GeV/$c$); solid and open black markers denote ALICE~\cite{Sahlmuller:2015mfq,Danisch:2021vmd} and PHENIX~\cite{PHENIX:2022qfp,PHENIX:2022rsx,PHENIX:2018che} data, respectively ($1.0<p_{\mathrm{T}}<5.0$ GeV/$c$). The pink line is the fit to the STAR data.}
\end{figure}

Figure~\ref{fig:DirectVirtualYield} shows the integrated direct virtual photon yields ($1<p_\text{T}<3$ GeV/$c$) as a function of $dN_\text{ch}/d\eta$ 
for Au+Au collisions at $\sqrt{s_\text{NN}}$ = 14.6, 19.6, 27, and 54.4 GeV. In current precision, the yields at a given \(dN_{\rm ch}/d\eta\) show no significant dependence on collision energy or centrality, indicating a common scaling behavior with charged-particle multiplicity. Together with the measurement at $\sqrt{s_\text{NN}}$ = 200 GeV~\cite{STAR:2016use}, the data are fitted using ${dN_{\gamma^{*}_\text{dir}}}/{dy} = A \times{(dN_\text{ch}/d\eta)}^{\alpha}$, with $A$ and $\alpha$ as free parameters. The fit gives $\alpha = $ 1.43 $\pm$ 0.04(stat.) $\pm$ 0.03(sys.) and $A = (8.49\pm1.64(\text{stat.})\pm2.00(\text{sys.}))\times10^{-5}$. 
The measured scaling exponent exceeds the expectations for both hadronic thermal radiation ($\alpha_\text{HG}\sim1.2$)~\cite{Shen:2013vja} and prompt photon production ($\alpha_\text{prompt}\sim1.2$), while remaining below the theoretical expectation for QGP thermal radiation ($\alpha_\text{QGP}\sim1.8$). This suggests that there is significant contribution from QGP phase. However, the discrepancy between STAR and PHENIX results in both magnitude and slope remains unresolved.


In general, this $dN_\text{ch}/d\eta$ dependence provides important insight into the underlying mechanisms of photon production in relativistic heavy-ion collisions. This contrasts with photons from initial hard scatterings in p+p collisions, which exhibit a much stronger dependence on beam energy~\cite{ALICE:2024vwy}. Since this $p_\text{T}$-integrated photon yield is from $p_\text{T}>1$ GeV/$c$, it is expected to predominantly reflect emission from a specific time window during the evolution of the QGP~\cite{Rapp:2014hha}.

\begin{figure*}
\includegraphics[width=0.80\textwidth]{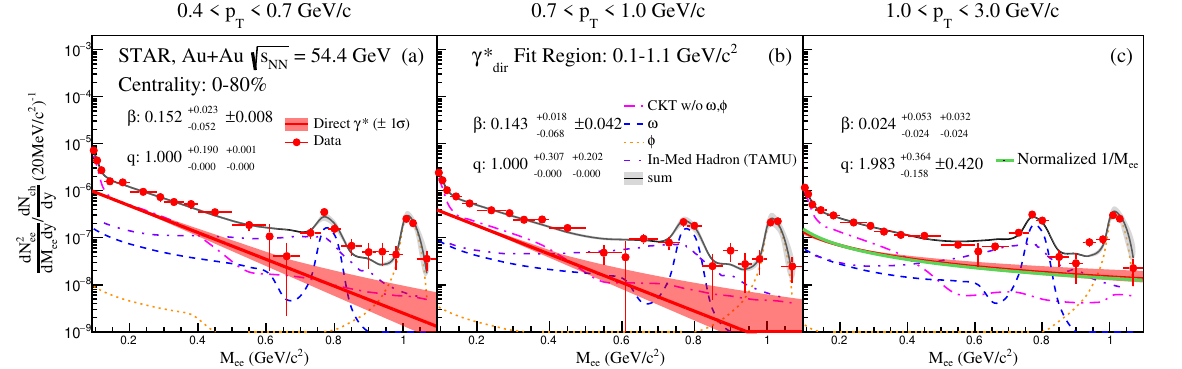}\\
\includegraphics[width=0.80\textwidth]{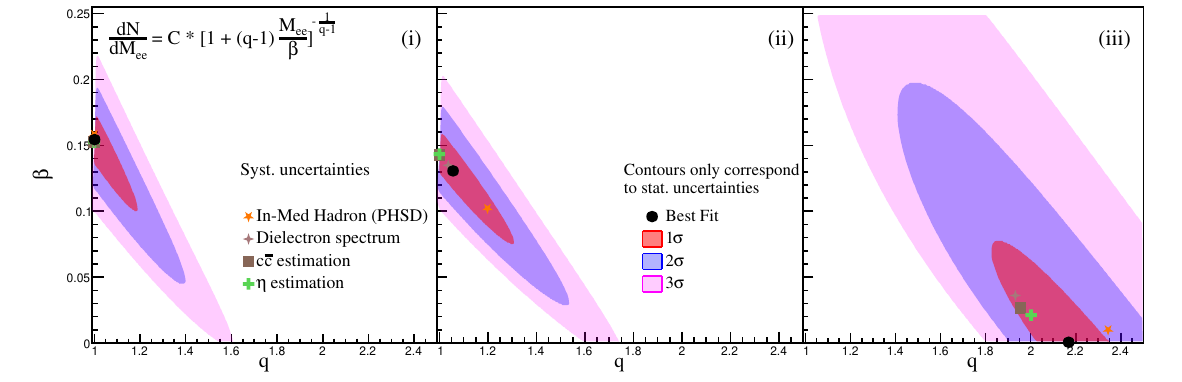}
\caption{\label{fig:DirVPMassDistribution}(Upper panels) Direct virtual photon mass distributions in Au+Au collisions at $\sqrt{s_\text{NN}}$  = 54.4 GeV in different $p_\text{T}$ ranges, for 0-80\% centrality. Red points represent the dielectron mass spectrum in different $p_\text{T}$ regions, red curves are the direct virtual photon fit and red band represents 1$\sigma$ band. (Lower panels) The constraints placed on virtual photon distributions, extracted from fits that incorporate scans in $q-\beta$ space. The red, blue and pink bands represent 1, 2 and 3$\sigma$ band respectively.}
\end{figure*}



After removing the hadronic cocktail from the inclusive dielectron spectrum, the direct virtual photon signal is expected to transition smoothly from thermal emission at low mass to hard scattering production at intermediate mass~\cite{STAR:2024bpc}. The excess dilepton contribution over $0.1<M_{ee} < 1.1\ \text{GeV/}c^{2}$ is extracted from 
\begin{equation}
    \label{eq:ThermalDielectron}
    \frac{dN^{\rm excess}}{dM_{ee}} = \frac{dN^\text{corr}_{ee}}{dM_{ee}}-\frac{dN_\text{CKT}}{dM_{ee}}, 
\end{equation}
where $\frac{dN^\text{corr}_{ee}}{dM_{ee}}$ and $\frac{dN_\text{CKT}}{dM_{ee}}$ represent the efficiency-corrected inclusive dielectron spectrum and the total expected background from known hadronic sources, respectively~\cite{Shao:2005iu,Han:2024nzr,STAR:2024bpc}. To study the direct virtual photon contribution and how the dielectron spectra evolve as a function of mass and $p_\text{T}$, the low mass ($0.1<M_{ee}<1.1\ \text{GeV}/c^{2}$) continuum must instead be fitted using a more detailed expansion due to contributions from the in-medium modifications to the $\rho$ spectral function. Hence, Eq.~\ref{eq:ThermalDielectron} expands to: 
\begin{equation}
    \label{eq:DirectVPMassDistribution}
\begin{split}
&\frac{dN_{ee}^\text{corr}}{dM_{ee}} = \gamma^{*}_\text{dir}+C_\text{CKT}^{w/o \ \omega,\phi} \times f_\text{CKT}^{w/o \ \omega,\phi}\\
&\quad\quad\quad\quad +C_{\omega}\times f_{\omega}+C_{\phi}\times f_{\phi}+H,
\end{split}
\end{equation}
where each term is defined below: 
\begin{itemize}
\item $f_{\omega}$, $\ f_{\phi}$, and $f_\text{CKT}^{w/o \ \omega,\phi}$ represent the mass shapes from $\omega$, $\phi$, and the remaining hadronic decays, respectively, with their corresponding amplitudes $\ C_{\omega}$, $\ C_{\phi}$ and $\ C_\text{CKT}^{w/o \ \omega,\phi}$. Due to the lack of $\omega$ and $\phi$ measurements, their amplitudes are left free in the fit.

\item $H$ accounts for the dielectron contribution from in-medium $\rho$-meson spectral function, estimated using the TAMU model ~\cite{vanHees:2006ng,Rapp:2014hha}. The systematic uncertainty therein is evaluated using the PHSD model~\cite{Cassing:2008sv,Cassing:2009vt,Cassing:1997jz}.

\item The direct virtual photon contribution ($\gamma^{*}_\text{dir}$) is modeled as
$$\gamma^{*}_\text{dir}=C\times f_\text{dir} = C \times  [1+(q-1)\frac{M_{ee}}{\beta}]^{-\frac{1}{q-1}}.$$
\end{itemize}
This parameterization can simultaneously describe exponential and power-law distributions, where parameter $q$ in $f_\text{dir}$ determines the shape of the virtual photon mass distribution~\cite{Shao:2009mu}. $\beta$ represents effective temperature when 
$q\simeq1$. Under this condition, the mass distribution approaches an exponential form, indicating the emission of direct virtual photons from a system close to thermal equilibrium. In contrast, values of $q$ closer to 2 result in power-law behavior ($1/M_{ee}$), characterized by the Kroll-Wada perturbative QED process~\cite{Kroll:1955zu,PHENIX:2008uif}.



The upper panels of Fig.~\ref{fig:DirVPMassDistribution} show the fit results of the direct virtual photon mass distributions in different $p_{\text{T}}$ intervals for Au+Au collisions at $\sqrt{s_{\text{NN}}} = 54.4$~GeV. The extracted parameter $q$ is consistent with $\sim1$ for $p_{\text{T}}<1\ \text{GeV}/c$, while it approaches $\sim2$ in the $1<p_\text{T}<3\ \text{GeV}/c$ interval, indicating a transition in the virtual photon mass distribution from an exponential shape to a power-law behavior. This change suggests that virtual photons at low $p_{\text{T}}$ originate from a system close to thermal equilibrium (i.e., the later stages of the collision), while those at higher $p_{\text{T}}$ are produced in more perturbative emission mechanisms (i.e., the earlier stages). 
The lower panels of Fig.~\ref{fig:DirVPMassDistribution} present comparisons between the dielectron spectra and the fit results across different $p_{\text{T}}$ intervals. Also shown are confidence level contours obtained from a $\chi^2$-minimization procedure~\cite{STAR:2018ldd,STAR:2019wlg}, based on a scan over the $q$ and $\beta$ parameter space. In particular, for the interval $0.4 < p_{\text{T}} < 0.7$~GeV/$c$, the extracted parameter is $q = 1.000^{+0.190}_{-0.000}\,^{+0.001}_{-0.000}$. In this case, the direct virtual photon distribution simplifies to an exponential form, $C \cdot e^{-M_{ee}/\beta}$, with an effective temperature of $\beta = 152^{+23}_{-17} \pm 8$~MeV. In this $p_{\text{T}}$ region, the blue shift effect will lower effective temperature due to the missing $e^+e^-$ pairs with higher mass. Considering this bias, this effective temperature is comparable with the one measured directly from the mass slope of the dielectron continuum in 0.4 to 1.2 GeV/$c^2$, which is $172\pm13\text{(stat.)}\pm\text{18(sys.)}$ MeV~\cite{STAR:2024bpc}. Additionally, in panel (c) of Fig.~\ref{fig:DirVPMassDistribution}, the green curve represents the direct virtual photon mass distribution extrapolated from our two-component fit to the lower mass range ($0.10 <M_{ee}<0.28$ GeV/$c^2$). 
This distribution is consistent with the extracted direct virtual photon component (red curve), confirming the internal consistency of the methods. Although the mass spectrum follows the perturbative QED Kroll–Wada form, the underlying source may still originate from thermalized $q\bar{q}$ pairs in the early stage, as suggested by the results in Fig.~\ref{fig:DirectVirtualYield}.

 The systematic uncertainties are evaluated by using PHSD instead of TAMU for the theoretical modeling of the in-medium $\rho^0$ modification, varying the dielectron spectrum, and the simulated cocktail background. The uncertainty of the theoretical model is the main contributor to uncertainty ($\sim20\%$) for $p_\text{T}>$ 0.7 GeV/$c$. The  other  sources of systematic uncertainty are shown in the lower panels of Fig.~\ref{fig:DirVPMassDistribution}. Similar transitions of $q$ are observed in all varied conditions.
 

\begin{figure}[b]
\includegraphics[width=0.47\textwidth]{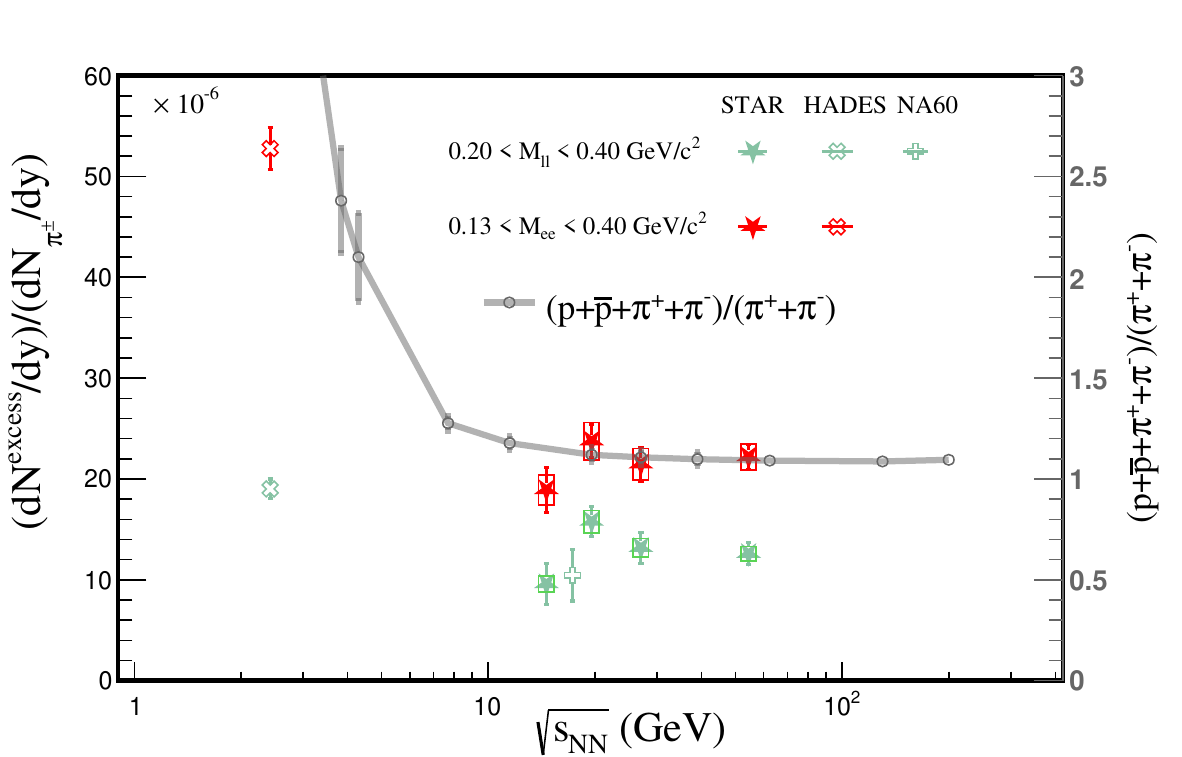}
\caption{\label{fig:ExcessYield} Excess dilepton yields normalized by $\pi^{\pm}$ yields in Au+Au collisions at $\sqrt{s_\text{NN}}$ = 14.6, 19.6, 27, and 54.4 GeV in 0-80\% centrality.
Green points represent dielectron excess yields within 0.2 $<M_{ee}< $ 0.4 GeV/$c^{2}$~\cite{NA60:2008dcb,HADES:2019auv}. Red markers represent dielectron excess yields within 0.13 $<M_{ee}<$ 0.40 GeV/$c^{2}$. Grey open circles represent ratio of $(p+\bar{p}+\pi^{+}+\pi^{-})/(\pi^{+}+\pi^{-})$~\cite{E-0895:2003oas,E895:2001zms,STAR:2017sal,STAR:2008med}.}
\end{figure}

Similar to the study of the scaling behavior of direct virtual photon production with charged-particle multiplicity across different beam energies, examining the dielectron yield as a function of beam energy may shed light on its production mechanisms and QGP evolution history~\cite{STAR:2015zal,STAR:2023wta}. Figure~\ref{fig:ExcessYield} presents the excess dilepton yields ($dN^\text{excess}/dy$) in the mass range $0.13 < M_{ee} < 0.40$~GeV/$c^{2}$, normalized by the charged pion yield, $(\pi^{+} + \pi^{-})/2$, at mid-rapidity~\cite{STAR:2019vcp,STAR:2017sal,Dhamija:2024cfh}, as a function of collision energy. The systematic uncertainties are dominated by the $\eta$ meson yield estimation (approximately 4--6\% depending on energy) and the dielectron spectra (also approximately 4--6\%). First, the STAR results at the same charged-particle multiplicity show no strong dependence on collision energy. The temperatures extracted from the low-mass region are found to be close to the critical temperature~\cite{STAR:2024bpc}, particularly at lower collision energies in the BES-II program. It should be noted that a clear difference is observed between the STAR and HADES excess yields in the region $0.13<M_{ee} <0.40$ GeV/$c^2$ ~\cite{STAR:2023wta,HADES:2019auv,HADES:2017jkt}. This is expected, as increased nuclear stopping at lower collision energies results in a higher baryon yield, thereby enhancing electromagnetic production from baryon-dominated processes such as $\Delta$ Dalitz decays, NN bremsstrahlung, and $\pi$N bremsstrahlung~\cite{HADES:2019auv}. The STAR results, which are in good agreement with the NA60 results, reflect the integrated yield in the mass range $0.20<M_{ll} <0.40$ GeV/$c^2$~\cite{NA60:2008dcb} and suggest a possible slight decrease from HADES~\cite{HADES:2019auv}. Finally, the observed variation in dilepton yields across different mass ranges at SIS energies (HADES) supports the interpretation that there is a strong decrease of baryon contribution at higher collision energies due to the exponential decrease of baryon transport~\cite{STAR:2024lvy}. This is qualitatively reflected in the $p/\pi$ ratio of $(p + \bar{p} + \pi^{+} + \pi^{-})/(\pi^{+} + \pi^{-})$~\cite{E-0895:2003oas,E895:2001zms,STAR:2017sal,STAR:2008med} shown in Fig.~\ref{fig:ExcessYield} grey markers, which serves as an indicator of baryon-to-pion production. At higher energies, the normalized dielectron yield is relatively consistent with the flat distribution of the $p/\pi$ ratio, with a possible increasing trend toward higher energy. It was shown in models~\cite{Rapp:2014hha} that such an increase may be correlated with the fireball expansion time. 

In conclusion, we report measurements of direct virtual photons in Au+Au collisions at $\sqrt{s_{\mathrm{NN}}}=14.6$--$54.4~\mathrm{GeV}$ using the internal-conversion technique. The virtual photon yield in $1 < p_\text{T} < 3~\mathrm{GeV}/c$ exhibits a robust power-law scaling with charged-particle multiplicity, with exponent $\alpha = 1.43 \pm 0.04~(\mathrm{stat.}) \pm 0.03~(\mathrm{sys.})$, and shows no discernible dependence on beam energy. The dielectron mass spectra reveal a continuous transition from an exponential form at low $p_\text{T}$, characterized by an effective temperature of $152^{+23}_{-17} \pm 8~\mathrm{MeV}$, to a power-law behavior at higher $p_\text{T}$, indicating a shift from thermal to perturbative emission mechanisms. The excess dilepton yields demonstrate that emission in the range $0.20 < M_{ll} < 0.40~\mathrm{GeV}/c^{2}$ and at higher BES-II energies is dominated by thermal radiation, whereas at lower masses ($0.13 < M_{ee} < 0.40~\mathrm{GeV}/c^{2}$) and at SIS energies, baryon-dominated processes provide the leading contribution to thermal dilepton production. This pattern mirrors the evolution of the baryon-to-pion ratio and reflects the diminishing role of transported baryons at higher collision energies.

We thank the RHIC Operations Group and SCDF at BNL, the NERSC Center at LBNL, and the Open Science Grid consortium for providing resources and support.  This work was supported in part by the Office of Nuclear Physics within the U.S. DOE Office of Science, the U.S. National Science Foundation, National Natural Science Foundation of China, Chinese Academy of Science, the Ministry of Science and Technology of China and the Chinese Ministry of Education, NSTC Taipei, the National Research Foundation of Korea, Czech Science Foundation and Ministry of Education, Youth and Sports of the Czech Republic, Hungarian National Research, Development and Innovation Office, New National Excellency Programme of the Hungarian Ministry of Human Capacities, Department of Atomic Energy and Department of Science and Technology of the Government of India, the National Science Centre and WUT ID-UB of Poland, German Bundesministerium f\"ur Bildung, Wissenschaft, Forschung and Technologie (BMBF),  Helmholtz Association, Ministry of Education, Culture, Sports, Science, and Technology (MEXT), Japan Society for the Promotion of Science (JSPS), and Agencia Nacional de Investigacion y Desarrollo de Chile (ANID), Chile.

\bibliography{./aps/apssamp.bib}

\end{document}


\title{Supplemental material: Probing the production mechanisms of virtual photon at RHIC}
\maketitle


\section{Detector and dataset descriptions}


The Au+Au collision data used in the direct virtual photon measurements at $\sqrt{s_{\mathrm{NN}}}$ = 14.6, 19.6, 27, and 54.4 GeV were collected with the Solenoidal Tracker at RHIC (STAR) detector~\cite{STAR:2002eio} during RHIC BES-II~\cite{Chen:2024aom}. Minimum-bias events were triggered by coincident signals in the Vertex Position Detector (VPD)~\cite{Llope:2014nva}.  Events are selected that have primary vertex coordinates along the beam line within $|V_{z}|<35$ cm and in the transverse plane from the beamline within $|V_{r}|< 2$ cm. After these event-selection criteria are applied, approximately 430, 250, 213, and 110 million events are retained for the analysis at $\sqrt{s_{\mathrm{NN}}}=$ 54.4, 27, 19.6, and 14.6 GeV, respectively. The collision centrality is determined based on the measured $dN_{\rm ch}/d\eta$ within $|\eta|<0.5$ and in comparison to a Monte Carlo Glauber simulation~\cite{Miller:2007ri}.
Charged-particle tracking and identification at midrapidity ($|\eta|<1$) are provided by the Time Projection Chamber (TPC)~\cite{Anderson:2003ur}, using momentum and $dE/dx$ measurements, together with the Time-of-Flight (TOF) detector~\cite{Llope:2012zz,Shao:2005iu}. TPC tracks are required to have at least 20 hit points for $\sqrt{s_\text{NN}}$ = 27 and 54.4 GeV, and 30 TPC hit points for $\sqrt{s_\text{NN}}$ = 14.6 and 19.6 GeV~\cite{Yang_2020}, with a distance of closest approach (DCA) to the primary vertex below 1 cm. A minimum number of TPC hits is also required for reliable $dE/dx$ particle identification. Electrons and positrons are identified using combined TPC and TOF information~\cite{Shao:2005iu,STAR:2024bpc,Han:2024nzr}, with $p_{\mathrm{T}}^{e}>0.2$ GeV/$c$ and $|\eta^{e}|<1$.


Opposite-sign ($+-$) and like-sign ($++/--$) electron pairs are reconstructed, with the latter used to estimate correlated and combinatorial backgrounds after correcting for acceptance differences between opposite and like-sign pairs using mixed events technique~\cite{STAR:2015tnn}. Additionally, the pairs from photon conversions in the detector material are removed by applying a cut based on their invariant mass and opening angle~\cite{STAR:2015tnn}. The dielectron spectra are corrected for detector efficiencies (TPC, TOF) and electron identification efficiency, binned in multiple dimensions ($\eta$, $p_\text{T}$, and centrality). The TPC tracking efficiency is determined using Monte Carlo embedding techniques, while TOF and other efficiency corrections are data driven~\cite{STAR:2015tnn}.

\section{Choice of invariant mass intervals}
Different invariant mass intervals are adopted in the present analysis because they are optimized for different physics objectives. The explanations for each mass interval are summarized below:

(a) $0.10<M_{ee}<0.28 \ \text{GeV/}c^{2}$: 
\begin{itemize}
    \item the Kroll-Wada formalism is applicable in the low-mass region ($M_{ee}\ll p_\text{T}$).
    \item the dominant $\pi^0$ Dalitz contribution is largely suppressed compared with the very low-mass region.
    \item the contribution from the in-medium $\rho$ spectral function is negligible in this mass range.
    \item the remaining dielectron spectrum is dominated by the hadronic cocktail (mainly contributed by $\eta$) and the direct virtual photon contribution, providing optimal sensitivity for the two-component fit.
\end{itemize}

(b) $0.10<M_{ee}<1.10 \ \text{GeV/}c^{2}$:
To investigate the invariant-mass distribution of direct virtual photons over an extended mass range. This analysis is performed only for the $\sqrt{s_{\rm NN}}$ = 54.4 GeV dataset because of its substantially higher statistical precision.
\begin{itemize}
    \item The lower mass bound (0.1 GeV/$c$) suppresses the region dominated by $\pi^{0}$ Dalitz decays and residual photon conversions.
    \item The upper mass bound (1.1 GeV/$c$) is selected to study the direct virtual photon mass distribution over the entire low-mass region, where the transition from thermal-like to perturbative production mechanisms is expected to occur.
\end{itemize}

(c) $0.13<M_{ee} < 0.40 \ \text{GeV/}c^{2}$:
\begin{itemize}
    \item  The published HADES excess dielectron spectra are reported starting from $M_{ee}=0.13\  \text{GeV/}c^2$. Using the same lower mass boundary allows a consistent comparison of the integrated excess yield between the HADES and STAR measurements.
    \item Above $0.40 \ \text{GeV/}c^{2}$, broadened $\rho$ contribution becomes increasingly important and would dominate the integrated excess yield instead of thermal radiation.
\end{itemize}

\section{Systematic uncertainty}
The systematic uncertainties in direct virtual photon measurement arise from the methodology of direct virtual photon extraction~\cite{STAR:2016use}, the estimation of the physics background, and the measurement of dielectron spectrum~\cite{STAR:2024bpc}. The following variations are used to evaluate these uncertainties:
\begin{itemize}
    \item Methodology of direct virtual photon extraction
    \begin{itemize}
        \item[-] Two-component fit mass range (GeV/$c^{2}$): $[0.10,0.28]$ (default), $[0.08,0.28]$, $[0.08,0.36]$, $[0.12,0.36]$.
        \item[-] Renormalization of mass range (GeV/$c^{2}$): $[0.00,0.03]$ (default), $[0.00,0.05]$
    \end{itemize}
    \item Estimation of the physics background
    \begin{itemize}
        \item[-] $\eta$ estimation: The high-$p_\text{T}$(5 GeV/$c$) $\eta/\pi^{0}$ ratio ($0.0470\pm0.017$) and propagate the resulting change to the extracted direct virtual photon yield.
        \item[-] $c\bar{c}$ estimation: The $c\bar{c}$ de-correlation effect is evaluated by varying randomly assigning the electron and positron azimuthal angles in the $c\bar{c}$ simulation from PYTHIA~\cite{STAR:2024bpc}.
    \end{itemize}
    \item Measurements of dielectron spectrum
    
    The overall systematic uncertainty which is $p_\text{T}$ independent, from the dielectron spectrum, labeled as "global", is dominated by track selection, and efficiency correction.
\end{itemize}

\begin{table}[htp]
\caption{Relative individual and  total systematic uncertainty for direct photon yields ($1<p_\text{T}<3\ \text{GeV/}c$) at 14.6, 19.6, 27, and 54.4 GeV.}
\label{table:SystUncertainty4Yield} 
\centering
\begin{tabular}{|c|c|c|c|c|}
\hline
\hline
Sources & 14.6 GeV & 19.6 GeV & 27 GeV  &54.4 GeV\\
\hline
 fit range       &   6.0\%    &9.3$-$14.8\%    &8.4$-$19.7\% &9.4$-$13.8\% \\
 normalization   &   0.8\%    &0.2$-$1.1\%     &0.8$-$2.5\%  &0.5$-$2.1\% \\
 $\eta/\pi^{0}$  &   4.3\%    &3.2$-$8.2\%     &5.6$-$20.9\% &4.3$-$19.1\% \\
 $c \bar{c}$     &   $<$ 0.1\%&$<$ 0.1\%       &0.2$-$0.4\%  &0.4$-$0.9\% \\
 global          &   7.7\%    &4.8\%           &7.6\%        &6.5\% \\
\hline
 sum             &   10.7\%   &10.9$-$17.6\%   &12.7$-$29.8\%&15.9$-$22.4\% \\
 \hline
\hline
\end{tabular}
\end{table}

\section{Direct virtual photon extraction}
\begin{figure}[b]
\includegraphics[width=0.45\textwidth]{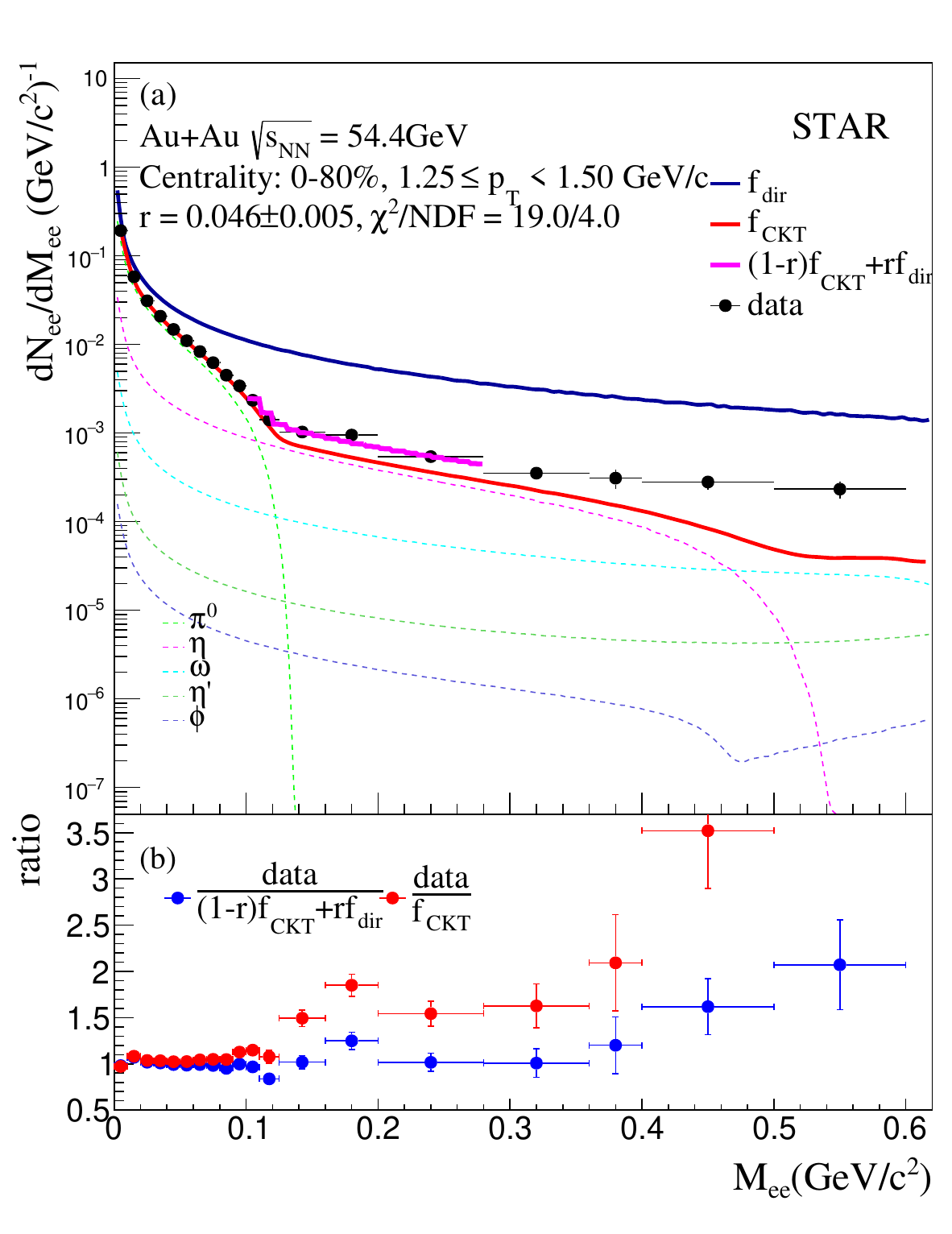}%
\caption{\label{fig:TwoComponentFit} Panel (a): The two-component fitting function results for 0-80$\%$ Au+Au dielectron spectra at $\sqrt{s_{\rm NN}}=$ 54.4 GeV, $1.25 < p_{\rm T}<1.50 \ \text{GeV}/c$ (black full circles). The magenta line is the fit to the data in the range $0.10<M_{ee}<0.28\ \text{GeV}/c^{2}$. Panel (b): Blue points represent the data divided by the fit model as a function of $M_{ee}$. Red points represent the data divided by the cocktail component as a function of $M_{ee}$.}
\end{figure}
The direct virtual photon contribution is extracted using a two-component fit~\cite{STAR:2016use} to the measured dielectron invariant-mass spectrum. The cocktail mass distribution $(f_{\rm CKT})$ is obtained from the hadronic decay simulation described above, while the direct virtual photon component $(f_{\rm dir})$ is determined from the internal-conversion relation~\cite{PHENIX:2009gyd}. Both components are normalized to the measured dielectron yield in the low-mass region dominated by $\pi^0$ Dalitz decays. The measured dielectron mass distribution is fitted with $(1-r) f_{\rm CKT}+rf_{\rm dir}$, where $r$ ($\gamma_{\rm dir}^{*}/\gamma_{\rm inclusive}^{}$) represents the direct virtual photon fraction. The fit is performed independently in each $p_{\rm T}$ and centrality interval. An example for 0-80$\%$ Au+Au collisions at $\sqrt{s_\text{NN}}$ = 54.4 GeV and $1.25 < p_{\rm T}<1.50 \ \text{GeV}/c$ is shown in Fig.~\ref{fig:TwoComponentFit}. The extracted fraction is used to determined the direct virtual photon yield.

\section{$\eta$ estimation}

The extraction of direct virtual photons via the internal conversion method relies on an accurate estimation of the hadronic cocktail, particularly the $\eta$ contribution in the low invariant mass region.

In previous measurements by the PHENIX Collaboration, the transverse momentum spectrum of the $\eta$ meson was estimated using $m_\text{T}$ scaling from the measured $\pi^0$ spectrum, assuming identical spectral shapes in transverse mass:

\begin{equation}
\left. \frac{dN_{\eta}}{dp_\text{T}} \right|{m_\text{T}} =
\left. \frac{dN{_{\pi^0}}}{dp_\text{T}} \right|{m_\text{T} \rightarrow \sqrt{p_\text{T}^2 + m{_\eta^2}}}.
\end{equation}

\begin{figure}[b]
\includegraphics[width=0.45\textwidth]{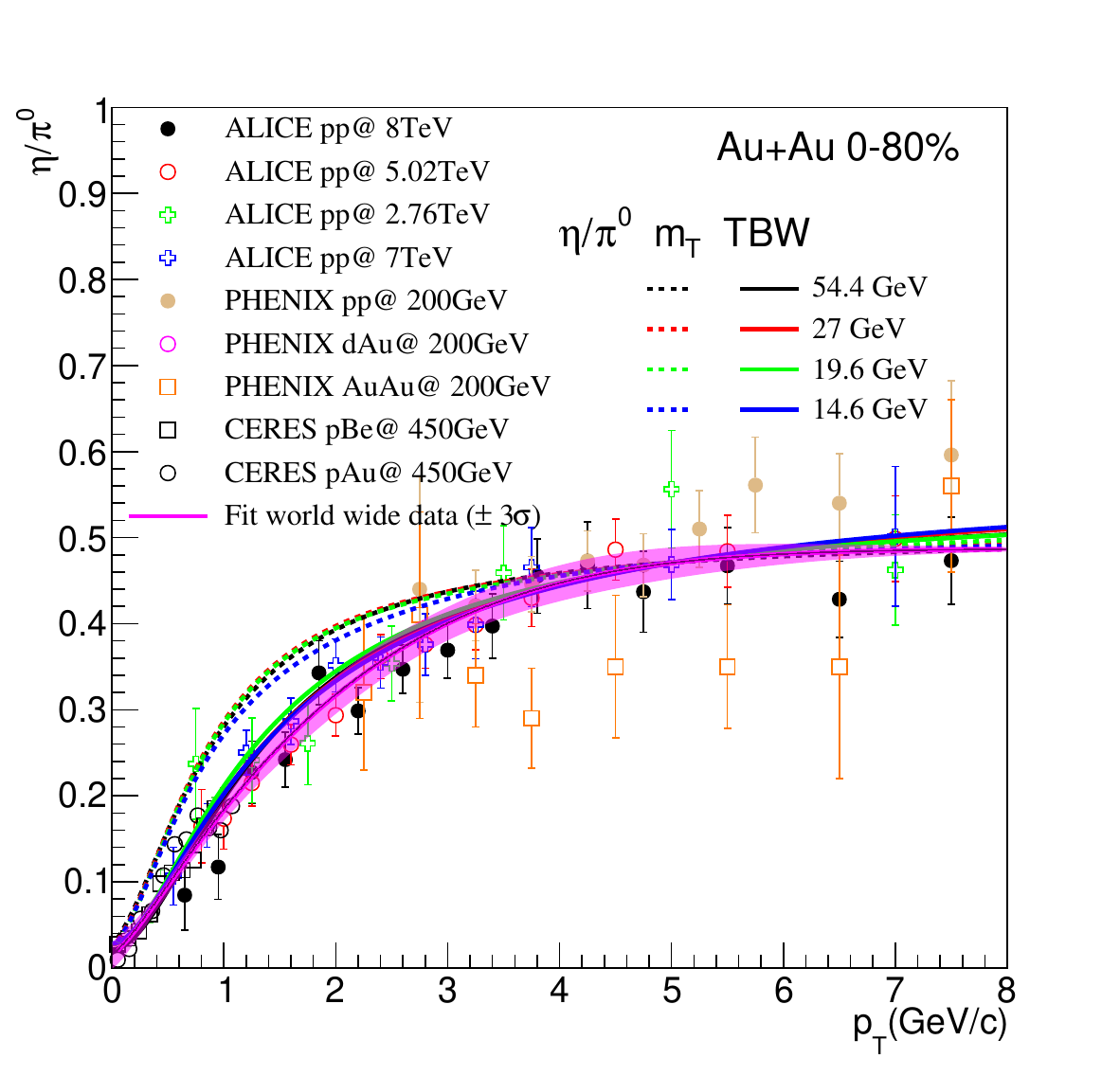}
\caption{\label{fig:Eta2Pi0_WWD} $\eta/\pi^{0}$ from $m_\text{T}$ scaling and TBW fit compared to $\eta/\pi^{0}$ from world wide data.}
\end{figure}

In contrast, the present STAR analysis employs a Tsallis Blast-Wave (TBW) parametrization constrained by simultaneous fits to identified hadron spectra, which incorporates collective flow effects. Table~\ref{table:TBWParaAuAu} presents the TBW parameters to predict each meson's $p_\text{T}$ shape in different centrality classes and collision energies. To quantify the impact of these assumptions, we generated $\eta$ spectra using both methods and evaluated their ratio as a function of $p_\text{T}$. The normalization of the $\eta$ meson is constrained using the high-$p_{\rm T}$ $\eta/\pi^0$ ratio extracted from a fit to the available world data. Compared with the compilation used in Ref.~\cite{Ren:2021pzi}, which yielded $\eta/\pi^0 = 0.487 \pm 0.024$, the present analysis includes additional PHENIX measurements in d+Au and Au+Au collisions~\cite{PHENIX:2006avo}. The updated fit, shown in Fig.~\ref{fig:Eta2Pi0_WWD}, results in $\eta/\pi^0 = 0.470 \pm 0.017$, which is adopted to normalize the TBW-predicted $\eta$ spectrum in this work. The corresponding normalization uncertainty is propagated over the full $p_\text{T}$ range through the TBW parametrization and included in the systematic uncertainty of the hadronic cocktail.

\begin{table*}[htp]
\caption{TBW parameters in different centrality classes Au+Au collisions at 54.4, 27, 19.6, and 14.6 GeV.}
\label{table:TBWParaAuAu} 
\centering
\begin{tabular}{|c|c|c|c|c|c|}
\hline
\hline
Energy (GeV) & Centrality (\%)  & q  &T (GeV)  &<$\beta$>\\
\hline
& 0-20      &   1.0442$\pm$0.0012   &  0.0965$\pm$0.0006   &  0.4348$\pm$0.0036\\
& 20-40     &   1.0580$\pm$0.0014   &  0.0979$\pm$0.0007   &  0.3726$\pm$0.0059\\
54.4 & 40-60&   1.0745$\pm$0.0014   &  0.0984$\pm$0.0007   &  0.2554$\pm$0.0110\\
& 60-80     &   1.0860$\pm$0.0006   &  0.0981$\pm$0.0006   &  0.0001$\pm$0.6398\\
& 0-80      &   1.0525$\pm$0.0008   &  0.0969$\pm$0.0004   &  0.3998$\pm$0.0029\\
\hline
 &0-20      &   1.0087$\pm$0.0040   &  0.1128$\pm$0.0018   &  0.4483$\pm$0.0057\\
 &20-40     &   1.0249$\pm$0.0045   &  0.1167$\pm$0.0023   &  0.3822$\pm$0.0095\\
 27 &40-60  &   1.0611$\pm$0.0022   &  0.1215$\pm$0.0026   &  0.0078$\pm$0.0081\\
 &60-80     &   1.0480$\pm$0.0021   &  0.1270$\pm$0.0025   &  0.0000$\pm$0.3923\\
 &0-80      &   1.0267$\pm$0.0050   &  0.1146$\pm$0.0030   &  0.3780$\pm$0.0100\\
\hline
 &0-20      &   1.0182$\pm$0.0060   &  0.1118$\pm$0.0028   &  0.4103$\pm$0.0113\\
 &20-40     &   1.0361$\pm$0.0058   &  0.1167$\pm$0.0029   &  0.3044$\pm$0.0194\\
 19.6&40-80 &   1.0613$\pm$0.0028   &  0.1134$\pm$0.0029   &  0.0001$\pm$0.5094\\
 &0-80      &   1.0380$\pm$0.0012   &  0.1100$\pm$0.0049   &  0.3400$\pm$0.0209\\
\hline
 14.6&0-80  &   1.0183$\pm$0.0072   &  0.1182$\pm$0.0033   &  0.3648$\pm$0.0182\\
\hline
\hline
\end{tabular}
\end{table*}


Importantly, the difference in the $\eta$ spectral shape results in a finite modification of the dielectron cocktail within the invariant-mass region used for the internal-conversion analysis. To quantify its impact on the direct virtual photon extraction, the fraction $r=\gamma_{\rm dir}^{*}/\gamma_{\rm inclusive}$ is determined using the two-component fitting procedure described above. The fits are performed independently in each $p_{\rm T}$ interval using the measured dielectron invariant-mass spectra in 0–80$\%$ Au+Au collisions at $\sqrt{s_{\rm NN}}=$14.6, 19.6, 27, and 54.4 GeV, together with the corresponding hadronic cocktails.

\begin{figure}[b]
\includegraphics[width=0.48\textwidth]{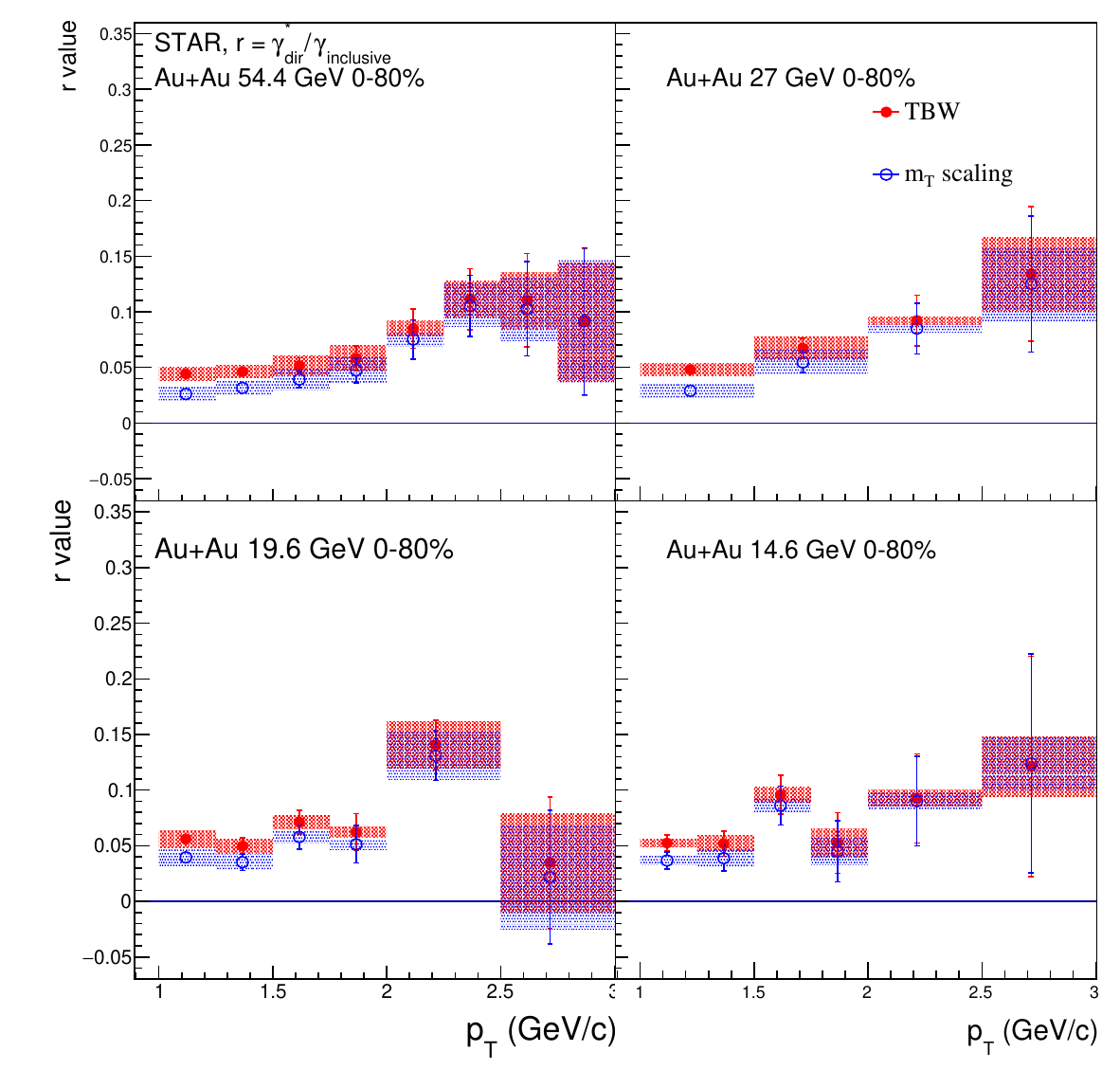}%
\caption{\label{fig:ParRComparsion} Direct virtual photon fraction as a function of $p_{\rm T}$ in 0–80$\%$ Au+Au collisions at $\sqrt{s_{\rm NN}}=$14.6, 19.6, 27, and 54.4 GeV. The fractions are determined from two-component fits to the measured dielectron invariant-mass spectra using the corresponding hadronic cocktails. Filled red circles denote the results obtained with the $\eta$-meson $p_{\rm T}$ spectra parameterized by the TBW model, while open blue circles denote those obtained using $m_{\rm T}$-scaled one. The vertical bars and boxes represent the statistical and systematic uncertainties, respectively. }
\end{figure}

The sensitivity to the $\eta$ contribution is evaluated by constructing the cocktail with two different descriptions of the $\eta$ $p_{\rm T}$ spectrum. In the default calculation, the $\eta$ spectrum is parameterized using the TBW function, whereas an $m_{\rm T}$-scaled spectrum is used as an alternative. The different spectral shapes and corresponding $\eta$ yields within the experimental acceptance modify the $\eta$ contribution to the dielectron cocktail. The two-component fit is repeated with each cocktail calculation, and the resulting $p_{\rm T}$-dependent direct virtual photon fractions are shown in Fig.~\ref{fig:ParRComparsion}.

The TBW parameterized $\eta$ spectrum yields systematically larger direct virtual photon fractions than those obtained with the $m_{\rm T}$-scaled spectrum. This comparison quantifies the sensitivity of the extracted direct virtual photon fraction to the modeling of the $\eta$ contribution, including both its $p_{\rm T}$ dependence and its yield within the experimental acceptance. Notably, the resulting variation is opposite in direction to the difference between the STAR and PHENIX direct virtual photon measurements~\cite{PHENIX:2022qfp,PHENIX:2022rsx,PHENIX:2018che}, indicating that the treatment of the $\eta$ spectrum alone does not account for the observed difference.

\section{Effect of the $p_\text{T}$ integration range}
The total yield is obtained by summing the STAR data points in the range $1 < p_\text{T} < 3$ GeV/$c$. To assess the effect of the different $p_\text{T}$ ranges, the Au+Au direct photon spectrum at 
$\sqrt{s_\text{NN}} = 200$ GeV measured by PHENIX~\cite{PHENIX:2008uif} 
is integrated over $1 < p_\text{T} < 3$ and $1 < p_\text{T} < 5$ GeV/$c$. 
The ratio of the two integrated yields, $\mathrm{Yield}_{1-3 \text{GeV}/c}/\mathrm{Yield}_{1-5\text{GeV}/c}$, is found to be 0.995, indicating 
that the discrepancy observed in Fig.~\ref{fig:DirectVirtualYield} is not driven by the difference in the $p_\text{T}$ integration ranges.

\section{Compatibility of the scaling behavior}
To further examine the scaling behavior shown in Fig.~\ref{fig:DirectVirtualYield}, additional fits were performed. The STAR data were fitted with the power-law form $A\times(dN_{\rm ch}/d\eta)^{\alpha}$, yielding $\alpha = 1.43$ with $\chi^{2}/{\rm NDF}=7.65/14$.
To test whether the PHENIX data are compatible with the same scaling exponent, the PHENIX data were fitted with $\alpha$ fixed to the STAR value while allowing only the normalization parameter $A$ to vary. This fit gives
$A=(2.57\pm0.06\text{(stat.)})\times10^{-4}$, but results in a poor fit quality of $\chi^{2}/{\rm NDF}=236.41/13$, indicating that the PHENIX data are not well described by the same slope as the STAR data. In contrast, a combined fit to the STAR and ALICE data with both $A$ and $\alpha$ free gives $A=(7.59\pm1.32\text{(stat.)})\times10^{-5}$ and $\alpha=1.45\pm0.03\text{(stat.)}$, with $\chi^{2}/{\rm NDF}=31.12/16$. The extracted exponent is consistent with the STAR-only result within uncertainties, suggesting that the STAR and ALICE data are compatible with a common scaling behavior.

\section{Scaling behavior of prompt photon contribution}

\begin{figure}[b]
\includegraphics[width=0.48\textwidth]{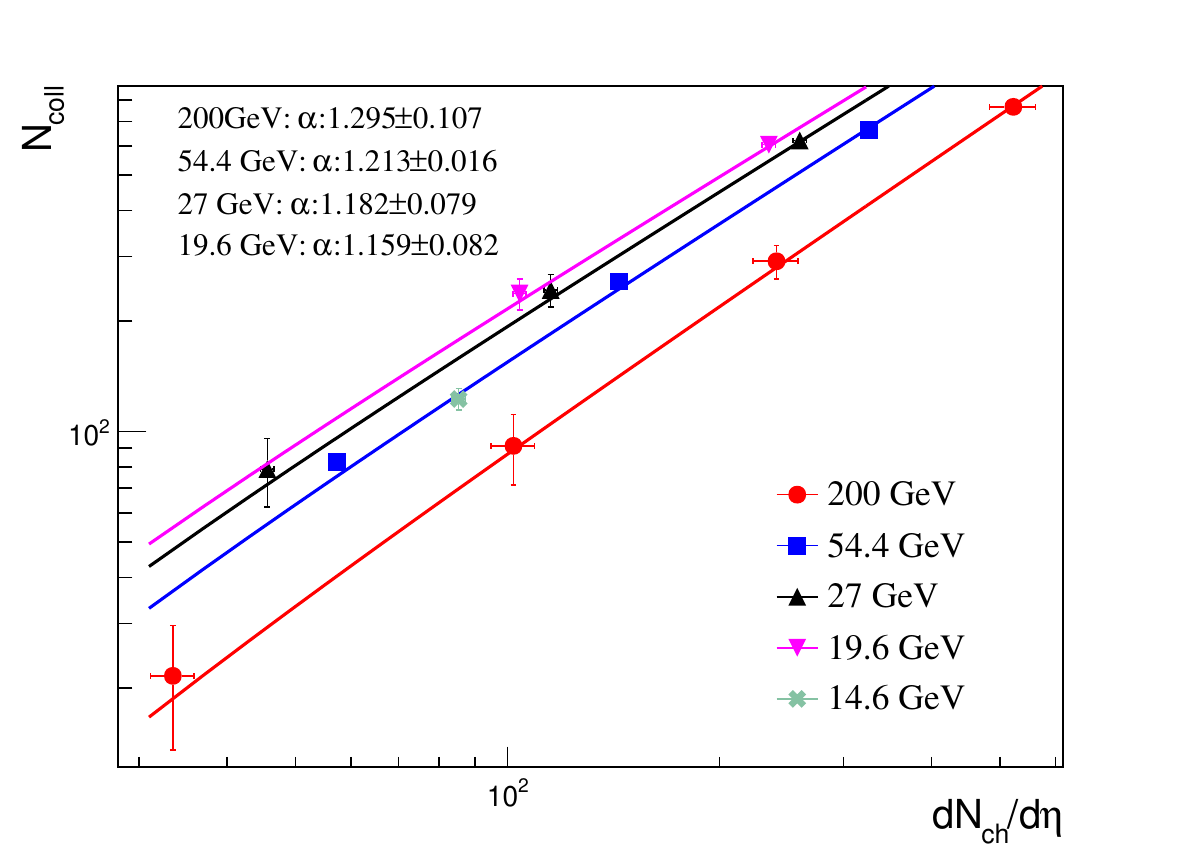}
\caption{\label{fig:NcollvsNch} $N_\text{coll}$ vs. $dN_\text{ch}/d\eta$ in Au+Au collisions at $\sqrt{s_\text{NN}}$ = 14.6, 19.6, 27, 54.4, and 200 GeV. The error bars on both $N_{\rm coll}$ and $dN_{\rm ch}/d\eta$ represent the quadratic sum of the statistical and systematic uncertainties.}
\end{figure}

Since the measured direct photon yield contains contributions from both prompt and thermal photons, it is important to estimate the multiplicity dependence expected for the prompt component. In heavy-ion collisions, prompt photon production is expected to follow binary-collision scaling and is therefore proportional to $N_\text{coll}$. Figure \ref{fig:NcollvsNch} shows the dependence of $N_\text{coll}$ on $dN_\text{ch}/d\eta$ for Au+Au collisions at $\sqrt{s_{\rm NN}}$ = 14.6, 19.6, 27, 54.4, and 200 GeV. 
The $N_\text{coll}$ values are obtained from the STAR Monte Carlo Glauber calculations~\cite{Miller:2007ri} corresponding to each collision energy and centrality class. Since $dN_\text{ch}/d\eta$ has not been published by STAR for BES energies, the values used in this work are estimated using the phenomenological parameterization described in Ref.~\cite{ALICE:2015juo,PHENIX:2015tbb}. 
The $\alpha$ values indicate that the prompt photon yield is expected to scale with charged-particle multiplicity with an exponent close to 1.2 over the studied beam-energy range. This behavior is comparable to the theoretical expectation for hadronic thermal radiation. Both are significantly weaker than the multiplicity dependence observed for the direct photon yield, for which $\alpha = 1.43 \pm 0.04\text{(stat.)} \pm 0.03\text{(sys.)}$. The stronger scaling observed in the data therefore cannot be explained solely by prompt photon production or hadronic radiation and is consistent with an additional contribution from thermal QGP radiation


\section{Hadronic cocktail normalization in the extended-mass fit}
\begin{table}[htp]
\caption{Normalization factors for the hadronic cocktail components obtained from the extended-mass fits to dielectron spectra in 0-80$\%$ Au+Au collisions at $\sqrt{s_{\rm NN}}=$54.4 GeV. The first and second uncertainties are statistical and systematic, respectively.}
\label{table:CocktailReNormalValue} 
\centering
\begin{tabular}{|c|c|c|c|c|}
\hline
\hline
$p_{\rm T}$ (GeV/$c$) & $C_{\rm CKT}^{w/o \ \omega,\phi}$  & $C_{\rm \omega}$  &$C_{\rm \phi}$    \\
\hline
 0.4-0.7     &   0.75$\pm$0.08$\pm$0.05   &  0.96$\pm$0.31$\pm$0.09   &  1.05$\pm$0.29$\pm$0.06 \\
 0.7-1.0     &   0.88$\pm$0.07$\pm$0.08   &  1.19$\pm$0.39$\pm$0.06   &  1.01$\pm$0.34$\pm$0.12 \\
 1.0-3.0     &   0.97$\pm$0.09$\pm$0.07   &  1.20$\pm$0.04$\pm$0.00   &  0.87$\pm$0.25$\pm$0.11 \\
\hline
\hline
\end{tabular}
\end{table}
In the extraction of the direct virtual photon mass distributions over the extended invariant-mass range, the contributions from the hadronic cocktail are simultaneously constrained by the measured dielectron spectra. The overall cocktail normalization, excluding the $\omega$ and $\phi$ contributions, is allowed to vary through a normalization factor $C_{\rm CKT}^{w/o \ \omega,\phi}$, while the $\omega$ and $\phi$ contributions are independently scaled by $C_{\omega}$ and $C_{\phi}$, respectively. Separating the $\omega$ and $\phi$ components prevents their localized contributions around the corresponding vector-meson masses from directly constraining the normalization of the remaining cocktail components.

The normalization factors are determined simultaneously with the parameters describing the direct virtual photon mass distribution in each $p_{\rm T}$ interval. Their extracted values for 0–80$\%$ Au+Au collisions at $\sqrt{s_{\rm NN}}=$54.4 GeV are summarized in Table~\ref{table:CocktailReNormalValue}. The first and second uncertainties represent the statistical and systematic uncertainties, respectively. The fitted normalization factors remain generally consistent with unity within uncertainties, indicating that only moderate adjustments of the nominal hadronic cocktail are required by the measured dielectron spectra.

\bibliography{./aps/apssamp.bib}